# Inhomogeneity Generated Waves in Pressure Balanced Structures

**Edisher Kaghashvili**

ekaghash@hotmail.com

**Inhomogeneity generated waves, discovered more than a decade ago, play an important role in processes like energy transfer, turbulence generation, heating, etc. To understand the nature of these waves we developed the formalism that looks for the short-term changes in the initial waveform due to the linear interaction of the initial disturbance and inhomogeneities in the background. Some aspects of inhomogeneity generated waves in the pressure balanced structures have already been studied in earlier papers, presented at the scientific gatherings, or proposed to different government funded agencies. Here, my intention is to demonstrate an existence of inhomogeneity generated waves in the pressure balanced structures in the simple analytical form that provide insight about them.**

In 2002, while doing my thesis about the wave process in solar plasma, I came across a very interesting phenomenon: when an initial (incident) wave interacts with an inhomogeneity driving waves that were not known or described in the literature. An unusual thing occurred in a simulation: one type of wave, specifically Alfvén wave named after its discoverer Hannes Alfvén (Ref. 1) generated waves of different frequencies when no mode coupling was expected according to existing studies of phase mixing and mode coupling. I considered a specific example related to inhomogeneous flow with constant shear. In that particular case, I could not only show new types of waves numerically, but I also characterized them spatially and temporarily, and obtained analytical solutions. These inhomogeneity generated waves were originally shown in MagnitoHydroDynamics (**MHD**), and later generalized for any physical system (Ref. 2-12).

In this letter, **I would like particularly emphasis one of its applications proposed early 2008. Specifically using this process as a diagnostics tool for waves detected in the solar higher layers. This topic was one of the main objectives of NASA's Solar Dynamic Observatory (SDO) mission.** Unlike standard seismology tools this method establishes a qualitative link between the detected compressive waves and the photospheric non-compressive driver waves capable of transporting energy long distance. The originally suggested mechanism operates the following way:

(a) Alfvénic (con-compressive) waves generated by interactions of the photospheric convective motions with footpoints of a magnetic tubes/loop propagate upwards along the magnetic field, and

(b) as they propagate, they interact with inhomogeneities of the medium through which they propagate producing inhomogeneity generated waves/fluctuations which are easier to detect because of their compressive nature.

Hence, this process unmasks the information about the driver's properties by providing information about photospheric spectrum capable of transferring energy to highest levels of the solar atmosphere. In what follows below, we will give solutions of the inhomogeneity generated waves in the pressure balanced structures that describe underlining mechanism applicable to numerous studies reported in the scientific literature in recent years. Namely, propagating initial wave interacting with inhomogeneities and exciting inhomogeneity generated waves capable of effectively exchanging energy with medium/plasma and particles.

Let us consider MHD system with isotropic pressure and without dissipation given by:

$$\frac{\partial \rho}{\partial t} + \nabla(\rho \boldsymbol{u}) = 0 \tag{1}$$

$$\rho \left[\frac{\partial \boldsymbol{u}}{\partial t} + (\boldsymbol{u} \cdot \nabla)\boldsymbol{u}\right] = -\nabla P + \frac{1}{4\pi}[\nabla \times \boldsymbol{B}] \times \boldsymbol{B} \tag{2}$$

$$\frac{\partial \boldsymbol{B}}{\partial t} + (\boldsymbol{u} \cdot \nabla)\boldsymbol{B} = (\boldsymbol{B} \cdot \nabla)\boldsymbol{u} - \boldsymbol{B}(\nabla \cdot \boldsymbol{u}) \tag{3}$$

$$\frac{\partial P}{\partial t} + (\boldsymbol{u} \cdot \nabla)P + \gamma P(\nabla \cdot \boldsymbol{u}) = 0 \tag{4}$$

$$\nabla \cdot \boldsymbol{B} = 0 \tag{5}$$

where $\rho$, $\boldsymbol{u}$, $\boldsymbol{B}$ and $P$ are density, flow, magnetic field and pressure.

We next perform a standard linearization procedure and represent physical variables as a sum of background state variable (using '0' subscript), and fluctuating variables representing linear waves:

$$\rho = \rho_0 + \rho' \tag{6}$$

$$\boldsymbol{u} = \boldsymbol{u}_0 + \boldsymbol{u}' \tag{7}$$

$$\boldsymbol{B} = \boldsymbol{B}_0 + \boldsymbol{b}' \tag{8}$$

$$P = P_0 + p' \tag{9}$$

The background inhomogeneous magnetic field is along x-axis, i.e. $\boldsymbol{B}_0 = (B_0, 0\ 0)$, where $B_0 = B_0(y)$, and there is no background flow: $\boldsymbol{u}_0 = 0$. The steady state equation that defines the pressure balanced structures is given by:

$$\frac{\partial}{\partial y}\left(P_0 + \frac{B_0^2}{8\pi}\right) = 0 \tag{10}$$

which states that variations in the background magnetic field and pressure along inhomogeneity axis (y-axis) are such that they maintain constant total pressure, which is a sum of the magnetic and thermal pressure.

Linear waves in this system are governed by the following system of equations:

$$\frac{\partial \rho'}{\partial t} + \rho_0 \left(\frac{\partial u'_x}{\partial x} + \frac{\partial u'_y}{\partial y} + \frac{\partial u'_z}{\partial z}\right) + u'_y \frac{\partial \rho_0}{\partial y} = 0 \tag{11}$$

$$\frac{\partial u'_x}{\partial t} = -\frac{1}{\rho_0}\frac{\partial p'}{\partial x} + \frac{b'_y}{4\pi\rho_0}\frac{\partial B_0}{\partial y} \tag{12}$$

$$\frac{\partial u'_y}{\partial t} = -\frac{1}{\rho_0}\frac{\partial p'}{\partial y} + \frac{B_0}{4\pi\rho_0}\left(\frac{\partial b'_y}{\partial x} - \frac{\partial b'_x}{\partial y}\right) - \frac{b'_x}{4\pi\rho_0}\frac{\partial B_0}{\partial y} \tag{13}$$

$$\frac{\partial u'_z}{\partial t} = -\frac{1}{\rho_0}\frac{\partial p'}{\partial z} + \frac{B_0}{4\pi\rho_0}\left(\frac{\partial b'_z}{\partial x} - \frac{\partial b'_x}{\partial z}\right) \tag{14}$$

$$\frac{\partial b'_x}{\partial t} + u'_y \frac{\partial B_0}{\partial y} = -B_0\left(\frac{\partial u'_y}{\partial y} + \frac{\partial u'_z}{\partial z}\right) \tag{15}$$

$$\frac{\partial b'_y}{\partial t} = B_0 \frac{\partial u'_y}{\partial x} \tag{16}$$

$$\frac{\partial b'_z}{\partial t} = B_0 \frac{\partial u'_z}{\partial x} \tag{17}$$

$$\frac{\partial p'}{\partial t} + u'_y \frac{\partial P_0}{\partial y} + \gamma P_0\left(\frac{\partial u'_x}{\partial x} + \frac{\partial u'_y}{\partial y} + \frac{\partial u'_z}{\partial z}\right) = 0 \tag{18}$$

$$\frac{\partial b'_x}{\partial x} + \frac{\partial b'_y}{\partial y} + \frac{\partial b'_z}{\partial z} = 0 \tag{19}$$

For simplicity we assume (a) isothermal process (temperature remains constant, i.e., $p' = c_s^2 \rho'$ where $c_s^2 = const.$ is a sound speed), and (b) no magnetic field fluctuations along the background magnetic field, $\boldsymbol{B}_0$. After some algebraic manipulation, we can derive two coupled equations for the velocity components along the magnetic field (longitudinal component) and along the spatial inhomogeneity (transverse component) given by:

$$\frac{\partial^2 u'_x}{\partial t^2} - c_s^2 \frac{\partial^2 u'_x}{\partial x^2} = -c_s^2 \left(\frac{1}{B_0}\frac{\partial B_0}{\partial y}\right)\frac{\partial u'_y}{\partial x} \qquad (20)$$

$$\frac{\partial^2 u'_y}{\partial t^2} - v_a^2 \frac{\partial^2 u'_y}{\partial x^2} - c_s^2 \frac{\partial^2 u'_x}{\partial x \partial y} = c_s^2 \left(\frac{1}{\rho_0}\frac{\partial \rho_0}{\partial y}\right)\frac{\partial u'_x}{\partial x} - \frac{c_s^2}{\rho_0}\frac{\partial}{\partial y}\left(\rho_0 \left[\frac{1}{B_0}\frac{\partial B_0}{\partial y} - \frac{1}{\rho_0}\frac{\partial \rho_0}{\partial y}\right] u'_y\right) \qquad (21)$$

where $v_a^2 = B_0^2/(4\pi\rho_0)$ is an Alfvén speed. Going forward, we will focus on inhomogeneity generated wave solutions for the coupled system (20)-(21).

Following the inhomogeneity generated wave formalism we can represent solutions as series in terms of two inhomogeneity parameters of the system

$$S_b \equiv \frac{1}{B_0}\frac{\partial B_0}{\partial y} = \frac{\partial \ln(B_0)}{\partial y} \qquad (22)$$

$$S_\rho \equiv \frac{1}{\rho_0}\frac{\partial \rho_0}{\partial y} = \frac{\partial \ln(\rho_0)}{\partial y} \qquad (23)$$

Alternatively, we can use the steady state equation (10) above that defines the pressure balanced structures and link inhomogeneity in background magnetic field, pressure/density and inhomogeneity in the Alfvén speed:

$$S_\rho = -\frac{2v_a^2}{2c_s^2+v_a^2}\frac{1}{v_a}\frac{\partial v_a}{\partial y} \equiv -\frac{2v_a^2}{2c_s^2+v_a^2}S_a \qquad (24)$$

$$S_b = \frac{2c_s^2}{2c_s^2+v_a^2}S_a \qquad (25)$$

where $S_a$ characterize inhomogeneity in the Alfvén speed.

Using above relationship, we can represent all background derivative terms in the equations with one inhomogeneity parameter only, and represent the solution as a series in terms of that parameter. Let us choose solutions as a series in terms of Alfvén speed inhomogeneity rate:

$$u_x(t,\mathbf{r}) = u_{x0}(t,\mathbf{r}) + S_a u_{x1}(t,\mathbf{r}) + S_a^2 u_{x2}(t,\mathbf{r}) + \ldots \qquad (26)$$

$$u_y(t,\mathbf{r}) = u_{y0}(t,\mathbf{r}) + S_a u_{y1}(t,\mathbf{r}) + S_a^2 u_{y2}(t,\mathbf{r}) + \ldots \qquad (27)$$

where zero-order terms (in $S_a$) are often given in the literature. **All other subsequent terms are explicitly $S_a$-parameter dependent and represent inhomogeneity generated waves.** In the inhomogeneity generated wave formalism, we first derive second order wave equations for the fluctuating variables, and substitute above expansions in these second order governing equations to derive solutions for waves. Afterwards, we group those equations according to their order in terms of the Alfven speed inhomogeneity parameter(s). **When the inhomogeneity parameter is small, obtained analytical solutions adequately describe inhomogeneity generated waves for the short times before more powerful nonlinear processes take over.**

The equations (20)-(21) are already the second order, and we can use (26)-(27) to derive solutions for linear waves. Note that the solutions for all other variables can be obtained from the solutions of $u'_x$ and $u'_y$. Here, we will not derive solutions for other components. **We only mention that it is straightforward to verify that, when no compressional waves are present in the system initially then all the density variations excited in the system later times are the inhomogeneity generated waves.**

Let us now solve our system using the inhomogeneity generated wave formalism. In the "zero" order ("0"-order solutions), we have

$$\frac{\partial^2 u'_{x0}}{\partial t^2} - c_s^2 \frac{\partial^2 u'_{x0}}{\partial x^2} = 0 \tag{28}$$

$$\frac{\partial^2 u'_{y0}}{\partial t^2} - v_a^2 \frac{\partial^2 u'_{y0}}{\partial x^2} - c_s^2 \frac{\partial^2 u'_{x0}}{\partial x \partial y} = 0 \tag{29}$$

According to the inhomogeneity generate wave formalism, these "0"-order waves in our system will interact with the non-uniform background and excite inhomogeneity generated waves. These two sets of equations can be solved for many possible initial conditions. We will consider a few important cases below.

As it was stated above, inhomogeneity generated waves in the pressure-balanced structures are generated when the initial "0"-order disturbance interacts with inhomogeneities in the medium through which it propagates. The system of equations for the "1$^{\text{st}}$-order" inhomogeneity generated waves is given by:

$$\frac{\partial^2 u'_{x1}}{\partial t^2} - c_s^2 \frac{\partial^2 u'_{x1}}{\partial x^2} = -\frac{2 c_s^4}{2 c_s^2 + v_a^2} \frac{\partial u'_{y0}}{\partial x} \tag{30}$$

$$\frac{\partial^2 u'_{y1}}{\partial t^2} - v_a^2 \frac{\partial^2 u'_{y1}}{\partial x^2} - c_s^2 \frac{\partial^2 u'_{x1}}{\partial x \partial y} = -\frac{2 c_s^2 v_a^2}{2 c_s^2 + v_a^2} \frac{\partial u'_{x0}}{\partial x} - \frac{2(c_s^2 + v_a^2) c_s^2}{2 c_s^2 + v_a^2} \frac{\partial u'_{y0}}{\partial y} - \frac{2(c_s^2 + v_a^2) c_s^2}{2 c_s^2 + v_a^2} A_1 u'_{y0} \tag{31}$$

where $A_1$ parameter is calculated from the inhomogeneity parameter derivative

$$\frac{\partial S_a}{\partial y} = A_1 S_a + A_2 S_a^2 + A_3 S_a^3 + \dots \tag{32}$$

The equation (32) is a general form (i.e. not all $A_i$ will be nonzero), and $A_i$-parameter calculation is straightforward after initial pressure-balanced structure is specified. The initial conditions for the "1"-order equations both the solutions and their derivatives are quiescent at $t = 0$. Hence we are interested in only inhomogeneous solutions.

The "2"-order system of equations is:

$$\frac{\partial^2 u'_{x2}}{\partial t^2} - c_s^2 \frac{\partial^2 u'_{x2}}{\partial x^2} = -\frac{2c_s^4}{2c_s^2+v_a^2} \frac{\partial u'_{y1}}{\partial x} \tag{33}$$

$$\frac{\partial^2 u'_{y2}}{\partial t^2} - v_a^2 \frac{\partial^2 u'_{y2}}{\partial x^2} - c_s^2 \frac{\partial^2 u'_{x2}}{\partial x \partial y} = -\frac{2c_s^2 v_a^2}{2c_s^2+v_a^2} \frac{\partial u'_{x1}}{\partial x} - \frac{2(c_s^2+v_a^2)c_s^2}{2c_s^2+v_a^2} \frac{\partial u'_{y1}}{\partial y} + \frac{4c_s^2 v_a^4}{(2c_s^2+v_a^2)^2} u'_{y0}$$

$$-\frac{2(c_s^2+v_a^2)c_s^2}{2c_s^2+v_a^2} A_2 u'_{y0} + -\frac{2(c_s^2+v_a^2)c_s^2}{2c_s^2+v_a^2} A_1 u'_{y1} \tag{34}$$

Here, again as in the "1"-order solutions, we are interested in the inhomogeneous solutions. After the problem and its solutions specified as in (20)-(21) and (26)-(27), it is straightforward to write governing equations in any order in terms of the inhomogeneity parameter.

**What determines how many terms to use in above series?**

This depends on the problem at hand. Since the solutions are $S_a$-parameter dependent, it will be the main factor. In the formalism, we are looking for the short-term changes in the initial waveform, and smaller $S_a$-parameter the better and longer our lower-order inhomogeneity generated wave solutions will be valid. As an example, let us calculate this parameter using conditions of one of the recent studies that examines the waves excited in the solar atmosphere (e.g., Ref. 14). As an example, the equilibrium density is assumed to be a function of y-axis only:

$$\rho(y) = \rho_{ext} + (\rho_{int} - \rho_{ext}) f(|y|) \tag{35}$$

where $\rho_{int}$ and $\rho_{ext}$ are the internal (near $y = 0$) and external/far from the origin density values, and the ratio is chosen, $\rho_{int}/\rho_{ext} = 5$. The function $f(|y|)$ is given by

$$f(|y|) = \frac{1}{1+(|y|/R)^\mu} \tag{36}$$

where $R = 1$ is a so-called mean-radius of density enhanced tube, and $\mu = 2.5$. The background magnetic field there is taken to be constant. Let us use above parameters, and estimate the inhomogeneity parameter. **Figure 1** shows the spatial variation of the inhomogeneity parameter $S_a$ in the perpendicular direction with respect to the background magnetic field, $\mathbf{B}_0$ for parameters shown above. The maximum magnitude $S_a = \pm 0.0432$ happens close to the mean-radius of density enhanced tube, $|y| \sim R$. **The smallness of this parameter guarantees that obtained analytical solutions adequately describe driven waves for the short times before more powerful nonlinear processes take over.** This is why, for most of the problems considered in the past, the inhomogeneity generated waves of lowest order were/are of interest. In addition, in most of the cases, these waves are dissipative and serve as a mean of energy redistribution though (a) classical dissipative terms in the equations, (b) inhomogeneity generated wave electric fields (e.g., Ref. 6), (c) various resonances processes (e.g., Ref. 3), etc.

## I. Specific Problems

Below we consider a few specific cases for above stated problem that are universal and have many applications.

### 1. Initial Alfvén wave propagating along the background magnetic field $B_0$

Let us investigate **one of the most interesting cases from the perspective of wave processes in the solar, space, magnetosphere and laboratory plasmas**. Specifically, when there is a wave in the system at $t = 0$ that is a most non-dissipative, long-lived and capable of propagating its energy long distances. In MHD such wave is Alfvén wave. In this particular case, for the initial Alfvén wave propagating, for example, along the background magnetic field, we have

$$u'_{x0}(t, \mathbf{r}) = 0 \tag{37}$$

$$u'_{y0}(t, \mathbf{r}) = u_{y0}(y) e^{i(k_x x + k_z z - \omega_a t)} \tag{38}$$

where $k_x$ and $k_z$ are the wave vectors along x- and z-axes. $\omega_a = k_x v_a$ is an Alfvén wave frequency. $u_{y0}(y)$ is a complex amplitude of the wave given by the initial conditions.

After initial drivers are specified, the governing equation for the longitudinal velocity fluctuation of the "1"-order inhomogeneity generated wave is

$$\frac{\partial^2 u'_{x1}}{\partial t^2} - c_s^2 \frac{\partial^2 u'_{x1}}{\partial x^2} = -\frac{2 i k_x c_s^4}{2 c_s^2 + v_a^2} u_{y0} e^{i(k_x x + k_z z - \omega_a t)} \tag{39}$$

The solution is given by:

$$u'_{x1}(t,\mathbf{r}) = \frac{ik_x c_s^4 u_{y0}}{2c_s^2+v_a^2} e^{ik_x x + k_z z} \left\{ \frac{2e^{-i\omega_a t}}{\omega_a^2 - k_x^2 c_s^2} + \frac{e^{ik_x c_s t}}{k_x c_s(\omega_a + k_x c_s)} - \frac{e^{-ik_x c_s t}}{k_x c_s(\omega_a - k_x c_s)} \right\} \quad (40)$$

As can be seen, the longitudinal component of the velocity fluctuation generated in the pressure-balanced structure has three propagating waves (1) Alfvén wave propagating along the x-axis, and (2) two waves of different amplitude propagating along and opposite background magnetic field $\mathbf{B}_0$ with sound speed.

The formal "1"-order solution of the transverse fluctuating component is given by

$$u'_{y1} = \int_0^t \sin(\omega_a[t-\tau]) \times \left\{ -\frac{2(c_s^2+v_a^2)c_s^2}{2c_s^2+v_a^2} \left( \frac{\partial u'_{y0}}{\partial y} + A_1 u'_{y0} \right) + ik_x c_s^2 \frac{\partial u'_{x1}}{\partial y} \right\} d\tau \quad (41)$$

and, one can continue so on and so on. Fortunately, it is not necessary as it was discussed above and has been shown in our early refereed publications (Ref. 2:13).

**Looking at the above solutions, we can see that the process operates as it was originally proposed in 2008.** There is a qualitative link between the compressive waves (longitudinal component $u'_x$ given by the equation (40) and closely related to compressibility of the wave) and the non-compressive driver wave capable of transporting energy long distance. Detection of the compressive waves is important for example in the solar observations where placing the instrument(s) locally difficult if not impossible. In such situations, the main methods of data collection are the imaging techniques.

**To leave no doubt that the same mechanism operates when different initial conditions are considered we will consider two addition cases.** Namely:

a) Initial driver wave by impulsive source, and
b) Spatially localized time-harmonic source.

Below, without loss of generality, we will drop "z"-coordinate dependence in the solutions, and add the source term that initiates the process. Here, our focus will be only to show that these two cases produce propagating wave along the magnetic field, which produces inhomogeneity generated waves as it interacts with inhomogeneities.

## 2. Initial driver wave by impulsive source

Let us consider the case of the initial pulse applied locally perpendicular to $\mathbf{B}_0$. This process was considered in numerous works were effects of such initial disturbances in different regions of the solar atmosphere were studied. The governing equations for the "0"-order driver wave are:

$$u'_x(t, \mathbf{r}) = 0 \tag{42}$$

$$\frac{\partial^2 u'_{y0}}{\partial t^2} - v_a^2 \frac{\partial^2 u'_{y0}}{\partial x^2} = u_m \delta(x)\delta(t) \tag{43}$$

where a wave source has magnitude $u_m$, and two delta functions showing the localization and impulsive nature of the initial source (localized on $t = 0$, and $x = 0$). The initial and boundary conditions are as follows:

$$u'_{y0}\big|_{t=0} = 0, \tag{44}$$

$$\frac{\partial u'_{y0}}{\partial t}\bigg|_{t=0} = 0 \tag{45}$$

$$u'_{y0}\big|_{x \to \pm\infty} = 0 \tag{46}$$

$$\frac{\partial u'_{y0}}{\partial x}\bigg|_{x \to \pm\infty} = 0 \tag{47}$$

A solution for the transverse velocity component is given by

$$u'_{y0}(t, \mathbf{r}) = u_{y0}(y) \times \frac{u_m}{2v_a} H(v_a t - |x|) \tag{48}$$

where $H$ is a Heaviside's unit step function. $u_{y0}(y)$ is the amplitude of the wave. The solution shows an expanding or outgoing wave along x-axis with uniform amplitude. It will drive the "1st-order" inhomogeneity generated waves according to the equations (30)-(31).

## 3. Spatially localized Time harmonic Source Perpendicular to $\mathbf{B}_0$

For this case, the governing equations for the "0"-order driver wave are

$$u'_x(t, \mathbf{r}) = 0 \tag{49}$$

$$\frac{\partial^2 u'_{y0}}{\partial t^2} - v_a^2 \frac{\partial^2 u'_{y0}}{\partial x^2} = u_m \delta(x) e^{i\omega_d t} \tag{50}$$

where a wave source has magnitude $u_m$, and $\omega_d$ is the frequency of the driver. Initial and boundary conditions are the same. The solution of the wave is given by

$$u'_{y0}(t, \mathbf{r}) = u_{y0}(y) \times \exp\left[i\omega_d \left(t - \frac{|x|}{v_a}\right)\right] \tag{51}$$

As can be seen, it is a solution of the spatially localized source propagating along x-axis in both positive and negative directions. In this case again, the driver wave given by (51) will drive the "1st-order" inhomogeneity generated waves according to the equations (30)-(31).

**To summarize**, we considered wave processes in the pressure-balanced structures. The specific examples we discussed are related to the wave processes in the solar atmosphere when the initial, long-lived waves generated near the solar surface propagate upward in the solar atmosphere and excites inhomogeneity generated waves along its path.

In a non-dissipative MHD system with anisotropic pressure, inhomogeneity generated waves can be generated by density, magnetic field, flow, pressure or temperature (both isotropic and anisotropic) spatial dependence as shown below by underlined terms:

$$\frac{\partial \rho}{\partial t} + (\boldsymbol{U}_0 \cdot \nabla)\rho + \underline{\rho(\nabla \cdot \boldsymbol{U}_0)} + \rho_0(\nabla \cdot \boldsymbol{u}) + \underline{(\boldsymbol{u} \cdot \nabla)\rho_0} = 0 \quad (52)$$

$$\rho_0 \left[\frac{\partial \boldsymbol{u}}{\partial t} + (\boldsymbol{U}_0 \cdot \nabla)\boldsymbol{u} + \underline{(\boldsymbol{u} \cdot \nabla)\boldsymbol{U}_0}\right] = -\nabla\left(P_0\left[\underline{\frac{\rho}{\rho_0} + \frac{T}{T_0}}\right]\right) + \frac{1}{4\pi}[\nabla \times \boldsymbol{b}] \times \boldsymbol{B}_0 + \underline{\frac{1}{4\pi}[\nabla \times \boldsymbol{B}_0] \times \boldsymbol{b}} \quad (53)$$

$$\frac{\partial \boldsymbol{b}}{\partial t} + (\boldsymbol{U}_0 \cdot \nabla)\boldsymbol{b} + \underline{(\boldsymbol{u} \cdot \nabla)\boldsymbol{B}_0} = (\boldsymbol{B}_0 \cdot \nabla)\boldsymbol{u} + \underline{(\boldsymbol{b} \cdot \nabla)\boldsymbol{U}_0} - \boldsymbol{B}_0(\nabla \cdot \boldsymbol{u}) - \underline{\boldsymbol{b}(\nabla \cdot \boldsymbol{B}_0)} \quad (54)$$

$$\frac{\partial p}{\partial t} + (\boldsymbol{U}_0 \cdot \nabla)p + \underline{(\boldsymbol{u} \cdot \nabla)P_0} + \gamma p_0(\nabla \cdot \boldsymbol{u}) + \underline{\gamma p(\nabla \cdot \boldsymbol{U}_0)} = 0 \quad (55)$$

where pressure is replaced by a product of temperature, $T$, and density in equation (53). Any initial disturbance propagating in this system is expected to be coupled with the inhomogeneity rates in underlined terms and excite linear inhomogeneity generated waves.

**FIGURE 1**: Spatial variation of the inhomogeneity parameter $S_a$ in the perpendicular direction with respect to the background magnetic field, $\mathbf{B}_0$. The profile is calculated using parameters from one of the cases from Ref 14.

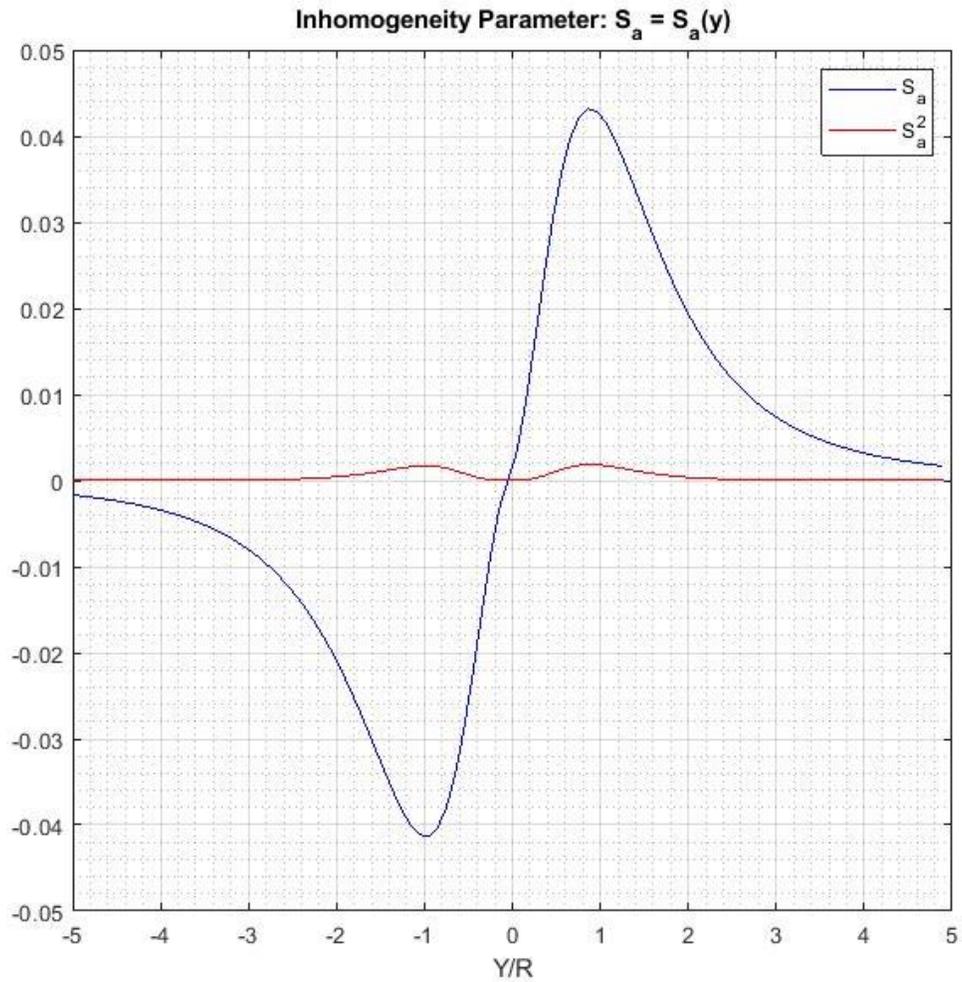